\documentclass[fleqn,usenatbib]{mnras}

\usepackage{newtxtext,newtxmath}
\usepackage[T1]{fontenc}
\usepackage{tabularx}
\DeclareRobustCommand{\VAN}[3]{#2}
\let\VANthebibliography\thebibliography
\def\thebibliography{\DeclareRobustCommand{\VAN}[3]{##3}\VANthebibliography}

\newcommand{\degree}{$^{\circ}$}

\newcommand{\jus} {JUS22\,}

\usepackage{graphicx}	
\usepackage{amsmath}	
\usepackage{multicol}

\usepackage[dvipsnames]{xcolor} 
\usepackage{bm}

\title[Super-Eddington accretion in slim disks]{Sustained super-Eddington accretion around neutron stars \& black holes}

\author[S. Ghodla \& J.J. Eldridge]{Sohan Ghodla$^{\thanks{sgho069@aucklanduni.ac.nz}{1}}$,  J. J. Eldridge$^{1}$ \\
$^{1}$Department of Physics, University of Auckland, Private Bag 92019, Auckland 1010, New Zealand}   

\date{Accepted XXX. Received YYY; in original form ZZZ}

\pubyear{2023}

\begin{document}
\label{firstpage}
\pagerange{\pageref{firstpage}--\pageref{lastpage}}
\maketitle


\begin{abstract}
   Recently, it was shown that the formation of a photon-trapping surface might not be sufficient to ensure unimpeded super-Eddington (SE) accretion. In light of this finding, here we derive a condition such that sustained and unimpeded SE accretion could be achieved in optically thick slim accretion disks surrounding neutron stars (NSs) and black holes (BHs). For this,  we calculate a semi-analytic approximation of the self-similar global radial velocity expression for an advection-dominated flow. Neglecting the influence of relativistic jets on the accretion flow, we find that for Eddington fraction $\Dot{m} \gtrsim 1.5 (\epsilon/0.1)^{3/5}$ ($\epsilon$ being the accretion efficiency) sustained SE accretion might be possible in slim disks around BHs irrespective of their spin. 
   The same condition holds for NSs when $\epsilon \gtrsim 0.03$. The presence of a surface magnetic field might truncate the disk at the magnetosphere of the NS, resulting in lower efficiencies and consequently changing the condition to $\dot{m} > 0.013 \epsilon^{-19/31}$. 
   Our approach suggests that sustained SE accretion might almost always be possible around NSs and BHs hosting accretion disks.


\end{abstract}




\section{Introduction}

It is understood that radiation takes longer to diffuse through an optically thick medium due to the presence of a higher number of scattering targets. Therefore, in the scenario where the medium is moving at a speed that is greater and in the opposite direction to the diffusive photons, the photons will be unable to overcome the flow \citep{Katz:1977}. Along similar lines, considering Bondi accretion \citep{Bondi_1952} onto a black hole (BH), \cite{Begelman_1978} calculated that for a given mass accretion rate $\Dot{M}$, a photon trapping surface would form at radius $r_{\rm tr}$  such that $ r_{\rm tr} =  2r_g \Dot{m}$
where 
\begin{equation}
    \Dot{m} := \epsilon \Dot{M}/\Dot{M}_{\rm Edd},
\end{equation}
 $\Dot{M}_{\rm Edd} = L_{\rm Edd}/ c^2$ being the mass accretion rate at Eddington luminosity, $L_{\rm Edd} = 4 \pi GM c/\kappa$. Since a trapping surface would form for any value of $\Dot{m} > 1$, an immediate conclusion is that steady-state super-Eddington (SE) Bondi accretion is always possible. \cite{Johnson_Sanderbeck_2022} (JUS22 from here on) found that the so-formed trapping surface would  ensure sustained and unimpeded SE accretion if and only if
\begin{equation}
    r_{\rm eq} < r_{\rm tr},
    \label{eq: Eqb radius vs trapping radius}
\end{equation}
where $r_{\rm eq}$ is the radius at which the momentum flux rate ($\Dot{p}_{\rm grav}$) of infalling matter in the gravitational potential is equal to the maximum momentum flux rate ($\Dot{p}_{\rm rad}$) of the outgoing radiation.
They derived a simple relation for the accretion rate
\begin{equation}
    \Dot{m}  \simeq 2/\epsilon
    \label{eq: jus condition}
\end{equation}
beyond which Eq. \ref{eq: Eqb radius vs trapping radius} is satisfied, hence allowing for unimpeded  Bondi accretion on a BH.
They further anticipated that the value of $\Dot{m}$ would be lower in the case of accretion through a disk.

While AGN observed today might be undergoing Bondi accretion owing to a low angular momentum (AM) supply of the gas (\jus), this might not the case for accretion in stellar binaries where the AM can be derived from the orbit. Therefore a disk is much more likely to form. Moreover, these disks would have outflows and be more luminous compared to those undergoing Bondi accretion, hence having observational significance. Therefore, it would be useful to extend the work of \jus for the case when accretion happens through a disk. Here we calculate the condition expressed in Eq. \ref{eq: jus condition} for such a form of accretion around neutron stars (NSs) and BHs while also accounting for the effect of mass outflow.

This article is organized as follows. In Section \ref{sec: extension of self similarity} we propose an extension of the radial velocity self-similar relation for advection-dominated accretion flow in slim disks around a Schwarzschild BH. In Section \ref{sec: condition for super-Edd} we use this to calculate a revised condition allowing for SE accretion in slim disks around NSs and BHs. Here we also examine the effect of NS's surface magnetic field on this condition. We end with a brief discussion in Section \ref{sec: discussion}.

\vspace{-0.5cm}
\section{Semi-Analytical approximation of globally self-similar radial velocity} \label{sec: extension of self similarity}

We assume a height-integrated, axis-symmetric and reflection-symmetric (w.r.t. the equatorial plane) steady state slim disk\footnote{As we are interested in accretion rates above Eddington, hence we choose geometrically slim rather than thin disks \citep{Abramowicz_1988}.}. In cylindrical coordinates $(r, \phi, z)$, this means all variables are only functions of radius $r$. General relativistic effects are included in a pseudo-Newtonian approach by employing the \cite{Paczynsky_Wiita_1980} potential
\begin{equation}
    \Phi = \frac{GM}{r - r_g},
\end{equation} 
 which gives a good representation of the relativistic potential of a Schwarzschild BH for $r \gtrsim 2r_g$. Here 
 \begin{equation}
     r_g = 2GM/c^2
 \end{equation}
 is the Schwarzschild radius of the central BH with mass $M$ and $c$ is the speed of light in vacuum, while the self-gravity of the accretion disk is neglected. In this potential, the  angular frequency $\Omega_{K}$ of a Keplerian circular orbit is given by
\begin{equation}
    \Omega_{K} = {\left(\frac{GM}{r^3} \right)}^{1/2} \frac{r}{r - r_g}.
    \label{eq: omega_K}
\end{equation}
The equation of local conservation of mass (in the disk) implies
\begin{equation}
    \Dot{M} = 2 \pi r \Sigma {\rm v}_r = 4 \pi r \rho H \mathrm{v}_{r},
    \label{eq: mass conservation}
\end{equation}
where $\Dot{M}, \Sigma, \rho, H, \mathrm{v}_{r}$ are the mass accretion rate, height integrated surface density, height averaged density, disk height from mid-plane and radial velocity, respectively. Considering the local conservation of angular momentum gives 
\begin{equation}
    \nu \Sigma  = \frac{\dot{M}}{3 \pi} \frac{f_*}{g},
\end{equation} 
where $f_*=1 -  \Omega_{\rm in} r_{\rm in}^2 / \Omega r^2$ and $g=-  \frac{2}{3} \frac{d \ln \Omega}{d \ln r}$. Here $\Omega$ is the angular frequency of the disk and $r_{\rm in}$ is the inner radius past which matter falls freely into the BH (e.g., \citealt{Frank_2002}).
Consequently, ${\rm v}_r$ takes the form
\begin{equation}
    {\rm v}_r = \frac{3 \nu}{2 r} \frac{g}{f_*} = \frac{3 \alpha \lambda^2 }{2}  \frac{g}{f_*} {\rm v}_K.
    \label{eq: radial velocity}
\end{equation} 
Here, $\nu = \alpha c_s H$ is the mean kinematic viscosity and is treated using the \cite{Shakura_Sunyaev_1973} approach. Additionally, for a disk in vertical hydrostatic equilibrium, $c_s = H \Omega_K$. We take $H = \lambda r$ and the expression for ${\rm v}_K$ follows from Eq. \ref{eq: omega_K}.
At the sonic radius $r_s$, we take ${\rm v}_r$ equal to the isentropic sound speed, i.e. (\citealt{Chen_1997, Narayan_1997})
%
\begin{equation}
    {\rm v}_r =  \lambda {\rm v}_K \sqrt{ \frac{2 \gamma^2} {\gamma+1}},
    \label{eq: velocity at sonic point}
\end{equation}
where $\gamma = 4/3$ is the ratio of specific heats for a radiation pressure dominated accretion flow. On the other hand, away from the BH ($r \sim 10^4 r_g$), Eq. \ref{eq: radial velocity} should converge to the self-similar solution of \cite{Narayan_Yi_1994} giving
\begin{equation}
{\rm v}_r = \frac{3 \alpha {\rm v}_K } {5+2 \epsilon^{\prime}} .
\label{eq: self similar}
\end{equation}
We take $\epsilon^{\prime} \simeq 1$, corresponding to a radiation pressure supported, strong advection-dominated accretion flow (ADAF).



To extend the form of ${\rm v}_r$ to an approximately global self-similar form, we need knowledge of $g/f_*$ in Eq. \ref{eq: radial velocity}.
Using Eq. \ref{eq: radial velocity}, \ref{eq: velocity at sonic point} and \ref{eq: self similar} we can (for a given value of $\lambda$) calculate $g/f_*$ near the two boundary points, i.e., at $r = r_s$ and $10^4r_g$.
In between, $g/f_*$ should be a function of $r$ and requires to be solved numerically. However, for a given value of $\lambda$ and $\alpha$, we find that $g/f_*$ can be reasonably approximated as (see Appendix \ref{sec: derive g/f_*})
%
\begin{equation}
    \frac{g}{f_*}  \simeq \left(\frac{r}{r_g }\right)^{\frac{1}{2} \left(a(\lambda) \left(\frac{r_g}{r}\right)^{3/2}  + b(\lambda)\right)}
    \label{eq: g/f_*},
\end{equation}
where 
\begin{equation}
\begin{aligned}
    & a(\lambda) = 7.70 \lambda^2 -17.38 \lambda + 34.41 + 7.15
\frac{{\rm log} 100 \alpha}{{\rm log (log}2)}, \\& 
    b(\lambda) = 0.76 \lambda^2  -1.72 \lambda + 0.74.
\end{aligned}
\end{equation}

\begin{figure}
    \centering
    \vspace{-0.6cm}
    \includegraphics[width = 1\linewidth]{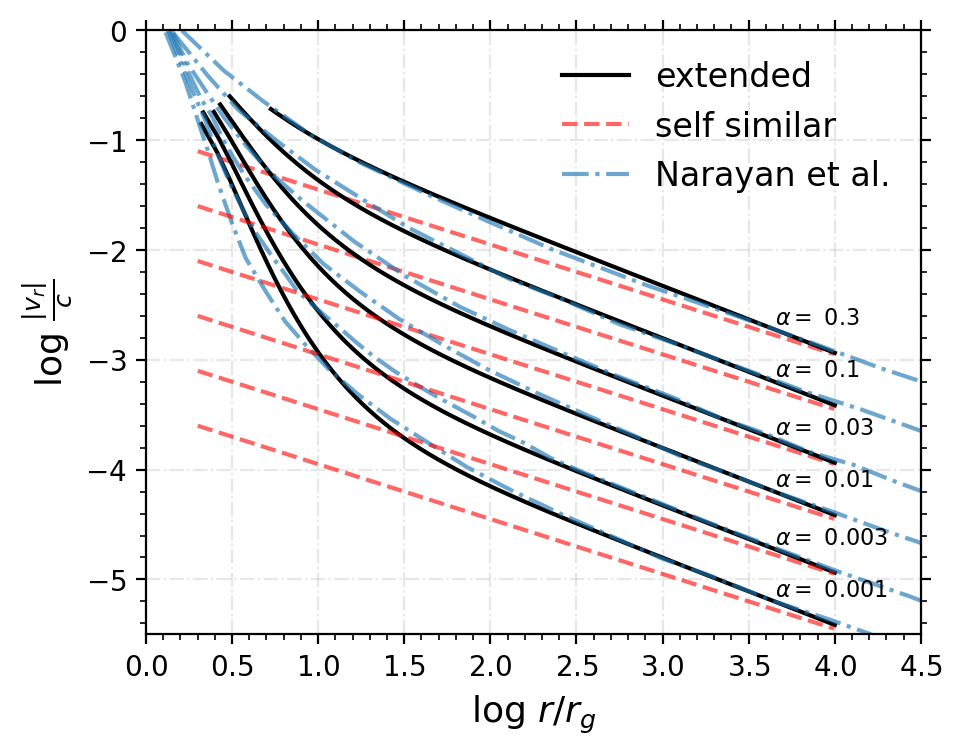}
    \vspace{-0.6cm}
    \caption{Extension of the self-similar solution for ${\rm v}_r$ in an ADAF around a Schwarzschild BH as discussed in Eq. \ref{eq: radial velocity} and \ref{eq: g/f_*}. Here we take $\lambda \equiv H/r = 1$. The cyan curves are (digitally extracted) numerical results from {\protect \cite{Narayan_1997}}. The red dashed lines correspond to Eq. \ref{eq: self similar}. The values of $r_s$ corresponding to different $\alpha$ parameters (in an increasing order) are: $r_s /r_g = 2.096, 2.152, 2.48, 2.7, 3.066, 5.315$ (source: \citealt{Narayan_1997, Popham_Gammie_1998}).}
    \label{fig: extended solution}
\end{figure}

 The resulting form of ${\rm v}_r$ is shown in Fig. \ref{fig: extended solution} and is in fair agreement to the numerically calculated global ${\rm v}_r$ solutions in \cite{Narayan_1997}. We note that for $r \lesssim 2r_g$, our calculation of ${\rm v}_r$ is imprecise as we assume $c_s \propto \Omega_K$ rather than $\Omega$. Moreover, ${\rm v}_r$ starts to diverge near $r_g$ in a pseudo-Newtonian potential (e.g., \citealt{Gammie_Popham_1998}). 

Since we are interested in accretion rates $\Dot{m} > 1$, under these circumstances, it is expected that the disk thickness parameter $\lambda \lesssim 1$. Moreover, the disk is expected to remain slim (i.e. $\lambda \lesssim 1$) for arbitrary large values of $\Dot{m}$ \citep{Lasota_2016}.
Lastly, for the current study, we adopt $\alpha \simeq 0.01$ corresponding to realistic optically thick and geometrically slim accretion disks in numerical simulations (e.g., \citealt{King_2007}).

\vspace{-0.3cm}
\section{Condition for super-Eddington accretion} \label{sec: condition for super-Edd}

\subsection{Effect of winds on mass accretion rate \& flow luminosity} \label{sec: Effect of mass outflow}

\begin{figure}
    \centering
    \vspace{-0.2cm}
    \includegraphics[width = 1\linewidth]{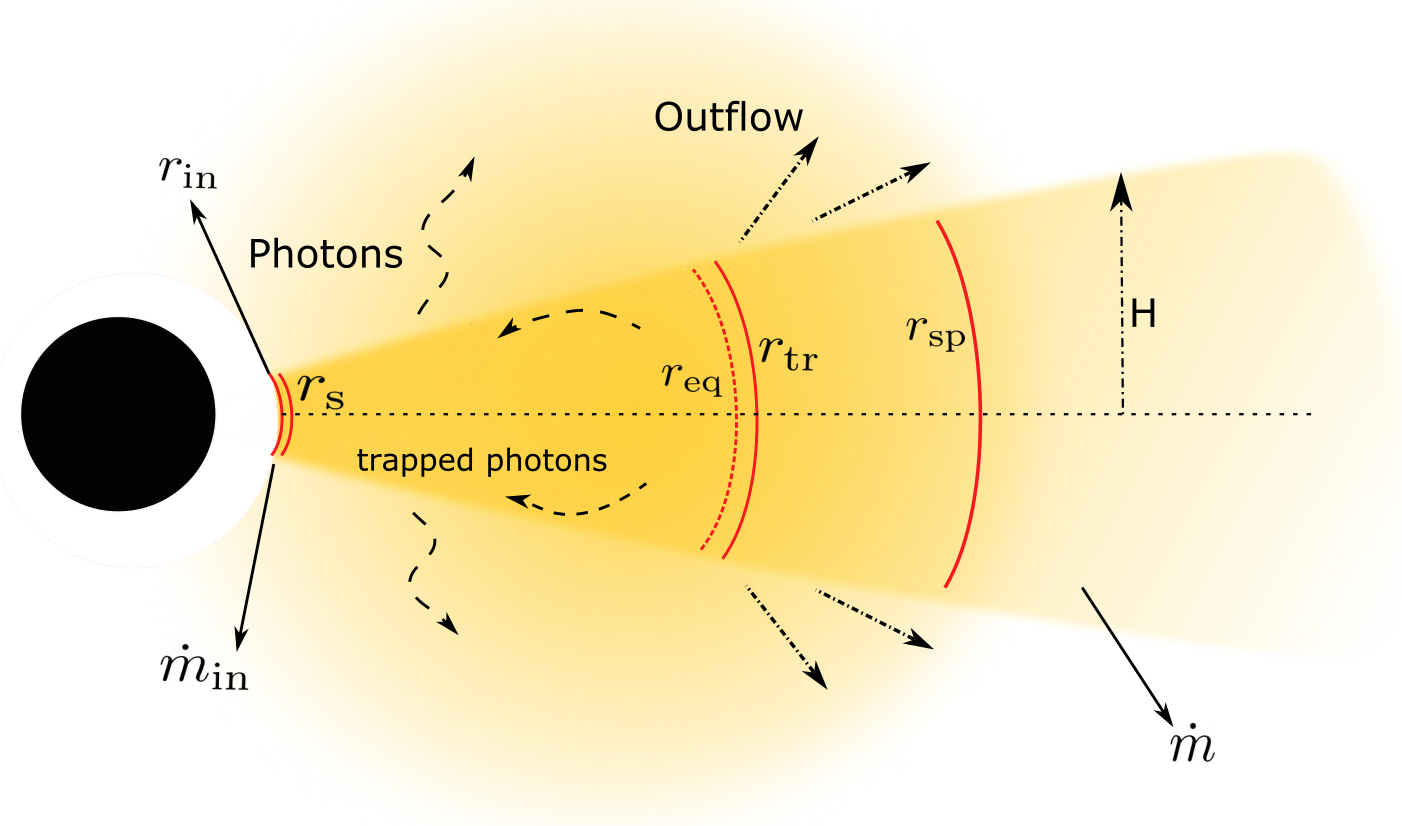}
    \vspace{-0.3cm}
    \caption{A schematic diagram of the accretion disk with the various processes and parameters considered here. Briefly, $r_{\rm in}, r_s, r_{\rm eq}, r_{\rm tr}, r_{\rm sp}$ are radii values corresponding to the inner disk, sonic point, momentum flux equilibrium, radiation trapping, and disk spherization respectively.}
    \label{fig: schematic}
\end{figure}

\jus assumed a steady state Bondi accretion. In practice for disk accretion, once $\Dot{m} \gtrsim 0.85$, mass outflows become inevitable (e.g., \citealt{McClintock_2006}). This is because, for accretion rates greater than this, the vertical pressure of gravity (even at its maximum value) could not  withstand the radiation pressure within the disk. 
The radius at which the local radiative flux generated by the disk's  viscosity equals the local Eddington flux is given by the spherization radius, $r_{\rm sp}$ (see Fig. \ref{fig: schematic} for a schematic).
Therefore, mass outflows are expected from within $r_{\rm sp}$. The fractional mass that reaches $r_{\rm in}$\footnote{The exact value of $r_{\rm in}$ is not important to us as most of this mass is expected to be lost in the region away from $r_{\rm in}$.} given a value of $\Dot{m}$ at $r_{\rm sp}$ is approximated as (for $\Dot{m} > 2.5$, see \citealt{Poutanen_2007})
\begin{equation}
    \frac{\dot{m}_{\mathrm{in}}}{\dot{m}} = \frac{1-a}{1-a\left(\frac{2}{5} \Dot{m} \right)^{-1 / 2}},
    \label{eq: outflow rate}
\end{equation} 
where $a=\epsilon_{\mathrm{w}}\left(0.83-0.25 \epsilon_{\mathrm{w}}\right)$ and $\epsilon_w$ is the efficiency at which the released energy goes into accelerating the outflow (i.e., not directly converted into radiation). We take $\epsilon_w = 1/2$. The mechanical feedback from the relativistic jets could act as another source of outflow. However, due to the bipolar nature of the jets, for the case of disk accretion, the effect of this feedback  would be substantially less compared to semi-Bondi accretion. Hence, for simplicity, here we ignore the impact of jets on the disk dynamics or the mass accretion rate.


Outside $r_{\rm sp}$ $(> r_{\rm tr}$), the disk luminosity $L = L_{\rm Edd}$ and the effective temperature follows the standard thin disk relation $T(r) \propto r^{-3/4}$. On the other hand, for $r \lesssim r_{\rm sp}$  the disk becomes radiatively inefficient, resulting in efficient photon trapping. Consequently, the radiation spectrum is that of a blackbody resulting in $T(r) \propto r^{-1/2}$.
Hence, for $r \leq r_{\rm sp}$ (assuming $r_{\rm sp} \sim r_{\rm tr}$)
\begin{equation}
    L \simeq 2 \int_{r_{\rm in }}^{r_{\rm tr}} \sigma T_{\rm eff}^4 2 \pi r \mathrm{~d} r \simeq L_{\rm Edd} \ln \left( \frac{r_{\rm tr}}{r_{\rm in}} \right) \simeq L_{\rm Edd} \ln \Dot{m}_{\rm in},
\end{equation}
where $r_{\rm tr}$ is taken from Eq. \ref{eq: r_tr vs mdot}. 
As such, the total disk luminosity (ignoring any extra contribution from jets) is
\begin{equation}
    L_{\rm tot} \simeq L_{\mathrm{Edd}}\left(1+\ln \dot{m}_{\rm in} \right).
    \label{eq: ADAF luminosity}
\end{equation}

\vspace{-0.4cm}
\subsection{Accretion flow around a Schwarzschild black hole} \label{sec: accretion flow around Schwarzschild BH}

Analogous to \jus, but with modifications to account for the effect of outflow and the nonspherical nature of inflow, we get
\begin{equation}
    \Dot{p}_{\rm grav}  = \gamma_{_L}  \frac{\Dot{M} {\rm v}_r}{4\pi r H},
    \label{eq: p_grav}
\end{equation}
where $\Dot{p}_{\rm grav}$ is the momentum flux rate of infalling matter and $\gamma_{_L} = 1/\sqrt{1 - {\rm v}_r^2 /c^2}$ is the Lorentz factor. However, due to outflows, only $\sim \Dot{M}_{\rm in}$ of the mass would be able to reach the inner region where most of the energy is released. Therefore, the resulting  maximum momentum flux rate of the outgoing radiation is
\begin{equation}
    \Dot{p}_{\rm rad}  = \frac{\epsilon \Dot{M}_{\rm in} c} {4\pi r^2},
    \label{eq: p_rad}
\end{equation}
where we assumed that the emission of radiation is isotropic.
Equating Eq. \ref{eq: p_grav} and \ref{eq: p_rad} and substituting for ${\rm v}_r$ gives
\begin{equation}
    \frac{3 \alpha \lambda \gamma_{_L}}{2^{3/2} \epsilon} \frac{\Dot{M}}{\Dot{M}_{\rm in}} \left(\frac{r}{r_g}\right)^{\frac{1}{2} \left(a(\lambda) \left(\frac{r_g}{r}\right)^{3/2}  + b(\lambda) -1 \right)} \frac{r}{r - r_g} = 1,
    \label{eq: solve for req}
\end{equation} 
where the form of $\Dot{M}/\Dot{M}_{\rm in}$ is given in Eq. \ref{eq: outflow rate} in a normalised fashion.

For $\epsilon_w = 1/2$ and $\Dot{m} = 10, 100, 1000$ we find $\Dot{m}_{\rm in} = 0.79\Dot{m}, 0.69\Dot{m}, 0.66\Dot{m}$ respectively.  To eliminate the dependence of $\Dot{M}/\Dot{M}_{\rm in}$ on $\Dot{M}$ in Eq. \ref{eq: solve for req}, we conservatively assume $\Dot{m}_{\rm in} = 2\Dot{m}/3$ for all $\Dot{m}$.  
The solution of Eq. \ref{eq: solve for req} then gives the radius at which the rate of momentum carried by the outgoing radiation equals that of the inwards accretion flow. The resulting solution is shown in Fig. \ref{fig: req solution} and approximately takes the form (valid for $\epsilon \gtrsim 0.03$)
\begin{equation}
      r_{\rm eq} \simeq \frac{(\lambda + 3.8) r_g}{(10\epsilon)^{2/5}}.
      \label{eq: req value}
\end{equation}
%



Radiation would be radially trapped if the characteristic radial diffusion timescale for photons $t_{\mathrm{diff}}$ is larger than the accretion timescale of the infalling matter $t_{\mathrm{acc}}$, i.e.,
\begin{equation}
    t_{\mathrm{diff}} / t_{\mathrm{acc}} \geq 1.
    \label{eq: diffusion timescale vs accretion time scale}
\end{equation}
Here $t_{\rm diff} = r^2/c \ell$; $\ell = m_{\rm p}/\sigma_{\rm T} \rho$ being the mean free path of the photons, $\sigma_{\rm T}$ being the Thompson scattering cross section assuming full ionisation and $t_{\rm acc} = r/{\rm v}_r$. On solving for the above condition, we get  (also see, \citealt{Ohsuga_2002})
\begin{equation}
  \frac{\Dot{m}_{\rm in} r_{g}}{2 \lambda \epsilon} \leq r_{\rm tr} <   \frac{\Dot{m} r_{g}}{2 \lambda \epsilon}.
     \label{eq: r_tr vs mdot}
\end{equation}
%
For $\alpha \simeq 0.01$ and any value of $\lambda$ (though $\lambda \simeq 1$ in the inner region)
Eq. \ref{eq: Eqb radius vs trapping radius} is satisfied for all $\Dot{m}_{\rm in} \gtrsim  (\epsilon/0.1)^{3/5}$. For $\Dot{m}_{\rm in} = 2\Dot{m}/3$, this gives
\begin{equation}
   \Dot{m} \gtrsim  \frac{3}{2} \left(\frac{\epsilon}{0.1}\right)^{3/5}.
    \label{eq: condition for f}
\end{equation}
For an accretion efficiency of $\epsilon = 0.1$, this gives $\dot{m} = 1.5$ compared to $\dot{m} = 20$ in Eq. \ref{eq: jus condition}. Therefore sustained SE accretion could be readily achieved (around accretion disks) as/once the flow becomes mildly SE.



\vspace{-0.3cm}
\subsection{Accretion flow around a Kerr black hole}

The above calculation has been conducted assuming that the accretion flow occurs around a non-rotating BH. In practice, as accretion proceeds, the BH should acquire AM. Does the above result hold for the subsequent accretion phase?

The spin of the BH would influence the form of Eq. \ref{eq: radial velocity}, hence affecting Eq. \ref{eq: req value}. Firstly, with increasing BH spin, numerical simulations show that $r_s$ recedes towards the horizon (e.g., \citealt{Popham_Gammie_1998}). Hence at a given radius near the horizon, the value of ${\rm v}_r$ for a non-rotating BH is larger than that for a Kerr BH in prograde motion w.r.t. the accretion disk. Additionally, $\epsilon$ would also become larger as $r_s$ (also $r_{\rm isco}$) decreases. Hence, a relatively smaller value of ${\rm v}_r$ at a given radii and/or a larger value of $\epsilon$ would move $r_{\rm eq}$ inwards. On the other hand, the condition for $r_{\rm tr}$ in Eq. \ref{eq: r_tr vs mdot} remains unchanged as it is implicitly independent of $\epsilon$. Therefore, the above conclusion of $\Dot{m} \gtrsim 1.5 (\epsilon/0.1)^{3/5}$ for sustained SE accretion remains valid for the case of a Kerr BH.

\vspace{-0.3cm}
\subsection{Accretion flow around a neutron star} \label{sec: flow around NSs}

In contrast to stellar BHs (of comparable mass), NSs have a larger radius and do not possess an event horizon. 
Considering a typical NS with radius $r_{\rm NS} =$ 12 km ($M_{\rm NS} =$ 1.4 M$_{\odot}$), this translates to $r_{\rm NS} \simeq 2.9 r_g$. For a given  $r_{\rm in}$, we calculate the corresponding value of $\epsilon$ as \citep{Novikov_Thorne_1973}
\begin{equation}
    \epsilon = 1-\sqrt{1- \frac{r_{\rm g }} {3 r_{\mathrm{{\rm in}}}}}.
    \label{eq: epsilon formula}
\end{equation}
As such for $r_{\rm in} = r_{\rm NS}$, this gives $\epsilon \sim 0.059$. 
Disregarding any change in the radius of the NS over the course of accretion, from Eq. \ref{eq: condition for f} this implies that sustained SE accretion could occur if $\Dot{m} \gtrsim 1.1$\footnote{Using Eq. \ref{eq: condition for f} for the case of a NS is justified as ${\rm v}_r$ (on which Eq. \ref{eq: condition for f} depends) should remain the same for a (non-rotating) NS as well due to Birkhoff's theorem.}. Due to the absence of a horizon, such accretion would release  energy at a tremendous rate of $\sim \epsilon \Dot{M}_{\rm in} c^2$ near the NS's surface, corresponding to a luminosity of $ L_{\rm tot} \simeq \Dot{m}_{\rm in} L_{\rm Edd} \simeq 1.76 \dot{m}_{\rm in} \times 10^{38}$ ergs/s. This could expel the accreted matter outwards to a mean relativistic velocity (calculated by setting ($\gamma_{_L} - 1)\Dot{M}_{\rm in} c^2 = \epsilon \epsilon_w \Dot{M}_{\rm in} c^2$):
\begin{equation}
    {\rm v} = c \sqrt{1 - \frac{1}{(\epsilon \epsilon_w + 1)^2}}.
\end{equation}
For $\epsilon = 0.059$ and $\epsilon_w = 1, 0.5, 0.25$ this results in v $\simeq 0.33c, 0.24c, 0.17c$ respectively. We note that the luminosity increases linearly with the mass accretion rate. This is in contrast to the $\ln{\Dot{m}_{\rm in}}$ rate of increase for the case of BHs in Section \ref{sec: Effect of mass outflow}.

\vspace{-0.3cm}
\subsection{Effect of the neutron star's surface magnetic field} \label{sec: Effect of magnetic field on accretion}

In practice, a NS is expected to have or develop (e.g., \citealt{NS_dynamo_2022}) a magnetic field over the course of its formation. This might then impede further accretion past the magnetosphere. Given a dipolar magnetic field 
\begin{equation}
    B \simeq \frac{\mu \sqrt{1+ 3 \cos^2 \theta}}{r^3}
\end{equation}
with a dipole moment $\mu$ and a dipole to the disk angle $\theta$ (which we adopt as $\theta = 90$\degree), we approximate the NS's magnetospheric radius as $r_m = \xi r_A$, where 
\begin{equation}
    r_A = \left [ \frac{\lambda \mu^2 \rm{v}_r} {2(\gamma_{_L} - 1) \Dot{M} c^2}  \right]^{1/4} 
    \label{eq: r_A}
\end{equation}
is the Alfvèn radius and is calculated by equating the energy density of $B$ with the infall kinetic energy density of the flow. 
Moreover, for an ADAF, it is expected that $\xi \sim 0.7-1$ \citep{Chashkina_2019}.

Now, if the magnetosphere radius $r_m$ of the NS is larger than its corotation radius (w.r.t the accretion disk), then the star is in the propeller regime where the accretion of matter past $r_m$ is prevented, hence imposing the condition $r_{\rm in} = r_m$ \citep{Illarionov_Sunyaev_1975}.

\begin{figure}
    \centering
    \vspace{-0.6cm}
    \includegraphics[width = 1\linewidth]{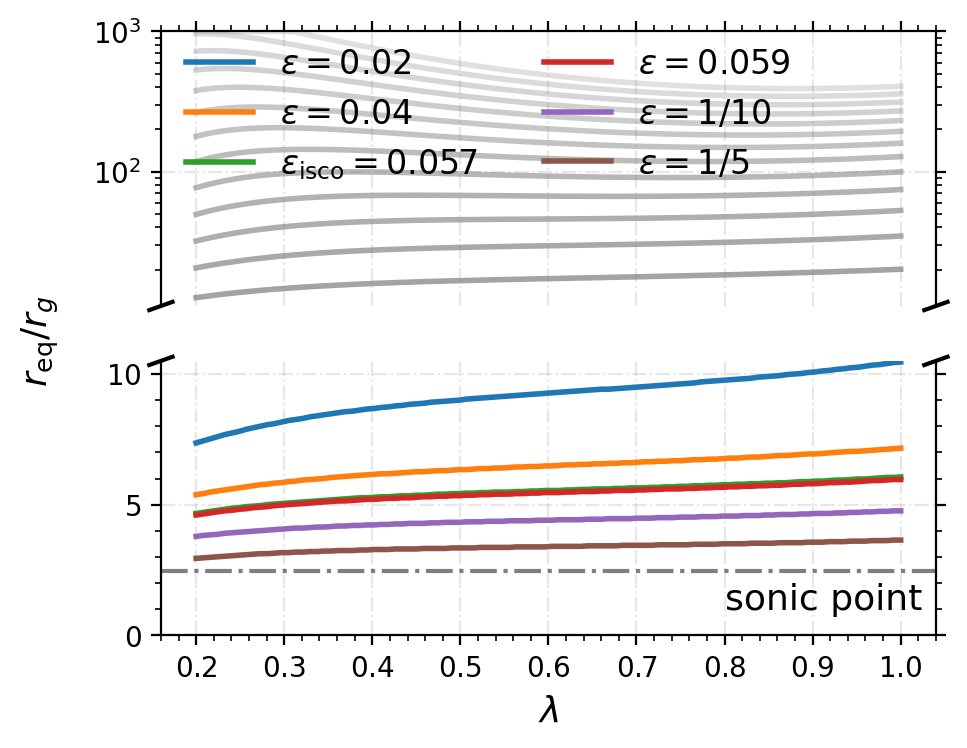}
    \vspace{-0.6cm}
    \caption{Figure shows the solution to Eq. \ref{eq: solve for req}, i.e., the radius $r_{\rm eq}$ at which the rate of momentum flux carried by the outgoing radiation equals that of the inwards accretion flow. Here we take $\alpha = 0.01$. The x-axis represents the vertical disk thickness parameter $\lambda \equiv H/r$. The $\epsilon$ values (in the legend) correspond to different $r_{\rm in}$ values. The grey curves range from $\epsilon = 0.03 - 0.0006$ each time $r_{\rm in}$ moving $15r_g$ away from the BH.
    The grey dashed-dotted line is the accretion flow's $r_s$ value. Note the change in scale on the y-axis.}
    \label{fig: req solution}
\end{figure}

Next, we aim to examine the effect of the propeller regime on the form of Eq. \ref{eq: condition for f}. Similar to the discussion in Section \ref{sec: accretion flow around Schwarzschild BH}, in the propeller regime, sustained SE accretion occurs up till $r_m$ only if both $r_m$ and $r_{\rm eq}$ are less than $r_{\rm tr}$. 
The form of $r_m$ as a function of $\epsilon$ can be calculated from Eq. \ref{eq: epsilon formula}, resulting in 
\begin{equation}
    r_m = \frac{r_g}{3 - 3(1 - \epsilon)^2}.
    \label{eq: r_m value}
\end{equation}
In addition, for $\lambda \simeq 1$ and $\epsilon < 0.03$, we find that $r_{\rm eq}$ (shown in Fig. \ref{fig: req solution}) could be approximated within a 10 per cent error margin as 
\begin{equation}
    r_{\rm eq} = (0.0043 \epsilon^{-50/31} + 8.5) r_g .
    \label{eq: r_eq at lambda = 1}
\end{equation}
As such, on comparing Eq. \ref{eq: r_m value} to Eq. \ref{eq: r_eq at lambda = 1} we can see that $r_{\rm eq} \gtrsim r_m$ for any given value of $\epsilon$. Therefore for sustained SE accretion, we only need to find the values of $\dot{m}$ such that the condition $r_{\rm eq} < r_{\rm tr}$ is also satisfied. For $\epsilon < 0.03$ and $\dot{m}_{\rm in} = 2/3 \dot{m}$ this results in 
\begin{equation}
    \dot{m} > 0.013 \epsilon^{-19/31}
    \label{eq: NS Super Eddington condition}
\end{equation}
and is depicted in Fig \ref{fig: Bsurf_rm_P} as a dotted red line. On the other hand, for $\epsilon \gtrsim 0.03$, the condition in Eq. \ref{eq: condition for f} takes over.
In Fig. \ref{fig: Bsurf_rm_P}, we also show the resulting $r_m$ values assuming that matter reaches $r_m$ at the reduced rate $\Dot{M}_{\rm in}$, i.e. replace $\Dot{M}$ with $\Dot{M}_{\rm in}$ in Eq. \ref{eq: r_A}.
The efficiency $\epsilon$ for these values of $r_m$ are plotted on RHS.
We further find that the magnetosphere radius for disk accretion $r_m \simeq 4 r_{m, \; {\rm Bondi}}$ where 
\begin{equation}
    r_{m, {\rm Bondi}} = \xi \left( \frac{\mu^2}{\Dot{M} \sqrt{2GM} }\right)^{2/7}
\end{equation}
and represents the magnetosphere radius for Bondi accretion. Here $r_m > r_{m, \; {\rm Bondi}}$ because at a given radius, the infall velocity v$_{r, \; {\rm disk}} < {\rm v}_{r, \; \rm Bondi}$. 

In Fig. \ref{fig: Bsurf_rm_P}, the region on the left of the grey dotted-dashed line is sub-Eddingtion while the one on the right has $\dot{m}_{\rm in} > 1$.
Only the pink shaded region satisfies Eq. \ref{eq: NS Super Eddington condition}, hence ensuring unimpeded SE accretion while the accretion is limited in the grey region.
This implies that a strong magnetic field ($B$  $\gtrsim 10^{12}$ G) impedes SE accretion (in the grey region) when the NS is in the propeller regime.
On the other hand, NSs with weaker magnetic field $B$  ($\lesssim 10^9$ G) or large spin period (so the NS is not in the propeller regime) can experience SE accretion at relatively low values of $\dot{M}$. The unshaded area lies in the sub-Eddington regime, where the accretion is not impeded by the luminosity.
We note that our values of $r_m$ are $\sim$ 6-8 times larger than those of \cite{Chashkina_2019} in the ADAF regime. Choosing $\lambda \simeq 1/2$ and matching our $\alpha$ to their choice, i.e., $\alpha = 0.1$, reduces these differences to within a factor of $\sim$ 2. 


\vspace{-0.5cm}
\section{Discussion} \label{sec: discussion} 

It has long been believed that SE accretion might occur at some phase of accretion onto compact remnants (aided by the formation of a photon-trapping surface). However, recently, \jus  found that the so-formed photon trapping surface would  ensure sustained and unimpeded SE accretion if and only if
$r_{\rm eq} < r_{\rm tr}$ (see, Eq. \ref{eq: Eqb radius vs trapping radius}).
Within the framework of our approach, we find that this condition can be readily fulfilled as/once the accretion flow becomes mildly SE. As a consequence, NSs and BHs accreting through disks could easily experience  SE accretion and sustain it to arbitrarily large values of accretion rate.  This then has implications for a variety of natural phenomena, some of which are discussed in Section  \ref{sec: implications}. But first, we present some of the caveats and uncertainties underlying the current work.

\subsection{Some caveats \& uncertainties}


1. \textit{Evidence for photon trapping}:
Photon trapping, which is crucial for the current work, has been demonstrated in 2D and 3D numerical simulations of slim disks undergoing SE accretion (e.g., \citealt{2005_Ohsuga, 2014_Sadowski, 2016_Sadowski}). Observationally, active SE accretion (for which photon trapping seems necessary) has been seen in some  Ultraluminous X-ray sources (see Section \ref{sec: ULXs discussion}).  \\

\noindent 2. \textit{Role of the $\alpha$ parameter}: 
An order of magnitude disagreement exists between observations and simulations regarding the true value of the disk viscosity parameter $\alpha$ (e.g., \citealt{King_2007}). 
Here we followed the numerical simulations and chose $\alpha = 0.01$. From Fig. \ref{fig: extended solution}, it can be seen that a larger $\alpha$ raises the value of $v_{r}$ for a given radius. Consequently, this raises the value of $r_{\rm eq}$ in Eq. \ref{eq: req value} hence requiring a larger value of $\Dot{m}$ to fulfill Eq. \ref{eq: Eqb radius vs trapping radius}. \\

\noindent 3. \textit{Disk Stability}: 
To allow for accretion over a range of $\Dot{m}$, we need to ensure that the disk remains stable.
The condition for thermal stability of the disk requires that $(d \ln Q^{+}_{\rm vis}/ d \ln H)_\Sigma < (d \ln Q^{-}/ d \ln H)_\Sigma $ 
where $Q^{+}_{\rm vis}$ is the rate of energy generation due to viscous heating and $Q^{-}$ is the rate of cooling due to inwards advection, radiative loss, and winds.
As such, the existence of disk wind would already have a stabilizing effect on the structure of the flow. 
In the inner region, Roche Lobe overflow and advection would also thermally stabilize the disk \citep{Abramowicz_1988}.

Subsequent studies have found slim disks also to be thermally stable against long wavelength perturbations \citep{Abramowicz_1995, Narayan_Yi_1995}.
Moreover, the slim disk branch is also viscously stable due to its positive slope on the S-curve. However, the transition process of the disk from a gas-dominated radiatively efficient flow at large radii to an optically thick ADAF near the BH remains uncertain. The work on the S-curve stability analysis shows that the transition region where the disk goes from a gas-to-radiation-dominated region is unstable, resulting in outbursts and limit-cycle behavior (e.g., \citealt{Czerny_2019}). The current work assumes that this transition is quick and lossless. Nevertheless, our results should remain valid given some mass reaches the spherization radius at a rate $\Dot{m}$.

\subsection{Implications} \label{sec: implications}


\subsubsection{Ultraluminous X-ray sources (ULXs)} \label{sec: ULXs discussion}

SE accretion on stellar BHs or NSs is one way of producing the apparently large luminosity ($L_X > 10^{39}$ ergs/s) observed in some ULXs (e.g., \citealt{2014_Bachetti, Israel_2017}). While accreting BHs tend to have a slower increase in the emitted luminosity as shown in Eq. \ref{eq: ADAF luminosity}, on the other hand, (see Section \ref{sec: flow around NSs}) for NSs, $L = \Dot{m}L_{\rm Edd}$.
This implies that a $\Dot{m} = 1000$ accretion on a BH would produce the same luminosity as $\Dot{m} = 8$ accretion onto the NS's surface. Moreover, to achieve a luminosity of $L > 10^{39}$ erg/s, a NS would only need to accrete at a rate $\dot{m} \simeq 6$.

\begin{figure}
    \vspace{-0.4cm}
    \centering
    \includegraphics[width = 1\linewidth]{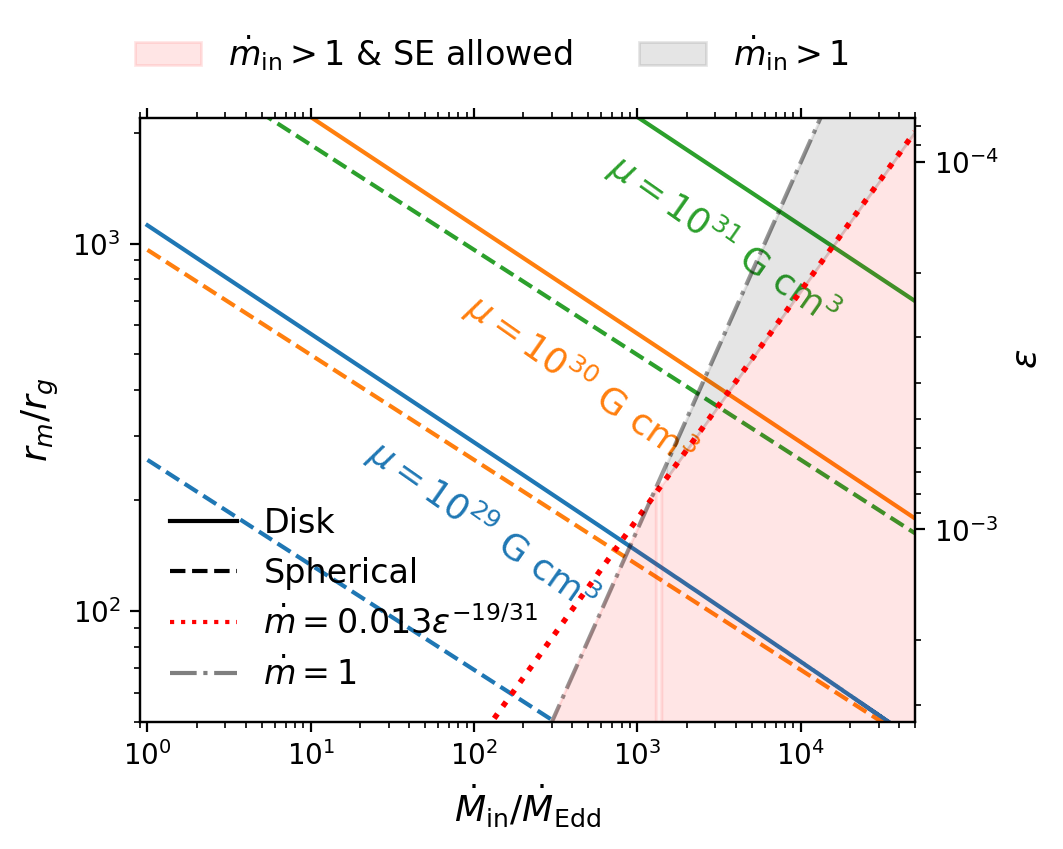}
    \vspace{-0.6cm}
    \caption{The magnetosphere radius (when $r_{\rm in} = r_m$) for various accretion rates and magnetic fields with the dashed and solid diagonal lines representing spherical and disk accretion, respectively. Here $\mu$ is the magnetic dipole moment of the NS, corresponding to surface magnetic field $B = 10^{11}, 10^{12}, 10^{13}$ G respectively, and we take $\xi = 0.8$. The RHS axis contains the efficiencies $\epsilon$ corresponding to the $r_m$ values on the LHS. The region on the left of the grey dotted-dashed line is sub-Eddington, while the one on the right has $\dot{m}_{\rm in} > 1$. For $\dot{m}_{\rm in} > 1$, only the pink shaded region satisfies Eq. \ref{eq: NS Super Eddington condition}, hence ensuring unimpeded SE accretion while the accretion is limited in the grey region.}
    \label{fig: Bsurf_rm_P}
\end{figure}

However, NSs hosting strong surface magnetic fields that have sufficiently small periods to be in the propeller regime would likely truncate the disk at $r_m$ (see Section \ref{sec: Effect of magnetic field on accretion}).
This then reduces the accretion efficiency, $\epsilon$, to low values, hence requiring significantly large $\dot{m}$ to produce high luminosity while some may be limited in their ability to under SE accretion (grey region in Fig. \ref{fig: Bsurf_rm_P}).

On the other hand, if the magnetosphere's barrier could be circumvented, then the accreted matter can be funneled along the magnetic axis of the NS. In this case, the photons would escape from the column’s side, and the radiation pressure would be inefficient in halting the inflowing matter \citep{2014_Bachetti}.  This, in conjunction with the relatively larger numerosity of lower mass stars\footnote{Stars with initial mass between $\sim$ 8-20 M$_{\odot}$ are more likely to form a NS with the BHs being produced at the higher masses. As such, owing to the nature of the initial mass function (e.g., \citealt{Kroupa2001}), more stars would end up producing a NS in comparison to a BH, with a suitable fraction of these NSs remaining bound to their binary companion post the Supernova explosion.} suggests that NSs might more frequently lie at the center of ULXs.


\vspace{-0.45cm}
\subsubsection{Rapidly spinning stellar mass black holes} \label{sec: implications: rapid spin}

Cores of massive stars might not retain a significant fraction of AM over the course of their evolution \citep{Fuller_2019, Eggenberger_2019}, hence producing BHs with low values of the dimensionless spin parameter ($a_* \lesssim 0.1$, \citealt{Fuller_Ma_2019}). On the other hand, some BHs have been observed in X-ray binaries to be rotating near critical (e.g., \citealt{ McClintock_2006, CygnusX1:Gou_2011}). 

At Eddington rate of accretion,  the BH mass growth rate (with initial mass $M_0$) takes the form: 
\begin{equation}
    M = M_{0}e^{ t/ (\epsilon t_{\rm Edd})}
\end{equation}
%
where $t_{\rm Edd} = M / \dot{M}_{\rm Edd} \simeq 0.45$ Gyr for electron scattering. Considering $\epsilon = 0.1$ for the entire course of accretion, this implies that for a BH to raise its mass by 25 per cent it needs to accrete at $\dot{M}_{\rm Edd}$ for $\sim 10$ Myr. This is larger than the lifetime of most massive companion stars ($M > 16$ M$_{\odot}$). Even a 25 per cent rise in the mass of the BH with initial $a_* = 0.1$ would increase its spin  to $a_* \sim 0.65$ \citep{Bardeen_1970, Thorne_1974}. Further growth of spin via accretion would occur at a monotonically decreasing rate, thus requiring relatively longer accretion timescales. Hence, some SE accretion episodes are needed to spin up the BHs to critical in a reasonable time period (e.g., \citealt{Sadowski_2011}). Our work suggests that such accretion could be readily achieved  even when the flow is mildly SE and sustained to arbitrarily high values.

\subsubsection{Supermassive BHs in active galactic nuclei (AGN)}

AGN, in particular, hosting quasars have been detected to redshifts $z \sim 7.5$ \citep{Banados}.  SE accretion is among the proposed scenarios to explain the masses of the supermassive BHs (SMBHs) residing at their centers (e.g., \citealt{Volonteri_2021}). However, observations have not yet conclusively found a SMBH accreting at such large rates (e.g., \citealt{2017_Trakhtenbrot}). The average spin parameter of SMBHs has been inferred to be $a \simeq 0.62$ (e.g., \citealt{2020_Jones}). Unless SMBHs are largely merger products of primordial/pop-III BHs that were themselves formed with significant AM, as discussed in Section \ref{sec: implications: rapid spin}, to gain $a_* \simeq 0.62$, these observed SMBHs should have accreted at least $\sim$ 25 per cent of their mass. This neglects the AM lost in jets or due to accretion from a retrograde flow. Assuming SMBHs arise from smaller seed BHs, this requires a significant mass accretion, mostly possible under SE scenario if detected at large redshifts.

Since the specific angular momentum necessary to form a Keplerian disk around a BH is proportional to $r_g$, the chances of semi-Bondi accretion increase for larger BH masses. Moreover, as $\Dot{M}_{\rm Edd}$ is proportional to the BH mass, a limited gas reservoir may also prevent SE accretion for larger masses. This suggests that SE accretion from disks is more likely to occur in galaxies hosting low-mass SMBHs, which are more likely to be found at larger, yet unobserved, redshifts. These disk-accreting SMBHs are expected to radiate at a luminosity given by Eq. \ref{eq: ADAF luminosity} and hence should be visibly SE. On the other hand, SMBHs experiencing SE Bondi accretion are expected to largely remain sub-Eddington (\jus). Overall, SMBHs accreting at a SE rate might transition from disk accretion at very high redshifts to some form of Bondi accretion at lower redshifts.

\vspace{-0.45cm}
\subsubsection{LIGO-Virgo-KAGRA (LVK) black hole mass spectrum}

Slowly rotating stars with pre-collapse, helium core mass between 65 - 135 M$_{\odot}$ are expected to end their life in a pair-instability supernova leaving behind no remnant \citep{Fowler_and_Hoyle1964}. This produces a mass gap in the BH mass spectrum resulting in no BHs with masses $\gtrsim 45$ M$_{\odot}$ (upto an upper limit, e.g., \citealt{Farmer_PISNe_2019}). As such, SE accretion onto less massive BH may explain the detection of some LVK BHs in the pair-instability mass gap \citep{van_son_2020}.

Moreover,  \cite{Briel_2022_b} showed that under the assumption of stable mass transfer in stellar binaries, the peak in the BH mass spectrum (around 35 M$_{\odot}$) as inferred by LVK might be reproduced by invoking unimpeded SE accreting BHs. The current analysis supports their work. We anticipate that allowing for wind outflows (which was not considered in \citealt{Briel_2022_b}) could have a dual effect. Firstly, the wind outflows during the SE phase would remove $\sim$ 10-30 per cent of the infalling mass, consequently reducing the mass of the BHs by a few per cent. Secondly, this makes the resulting BHs less massive and hence harder to merge (via the emission of gravitational radiation), consequently causing some of them to disappear from the peak. Cumulatively, we anticipate the peak to become less profound and shift to the left. \\

To summarise, we suggest that sustained SE might almost always be possible around NSs and BHs hosting accretion disks unless the NS is in the propeller regime and has a large surface magnetic field. In this case, SE accretion becomes harder to achieve. We briefly discussed the implication of our result for ULXs, BH spin, SMBHs mass in AGN, and the LVK BH mass spectrum.



\vspace{-0.6cm}
\section*{Acknowledgements}
We thank Dr. Maciek Wielgus for helpful comments on the manuscript. SG is supported by the University of Auckland doctoral scholarship. SG is grateful to John Bray for feedback. JJE acknowledge support of Marsden Fund Council managed through Royal Society Te Apārangi.

\vspace{-0.6cm}
\section*{Data Availability}

No dataset was generated in this work.



\vspace{-0.6cm}
\bibliographystyle{mnras}
\bibliography{refs} 

\begin{thebibliography}{}
\makeatletter
\relax
\def\mn@urlcharsother{\let\do\@makeother \do\$\do\&\do\#\do\^\do\_\do\%\do\~}
\def\mn@doi{\begingroup\mn@urlcharsother \@ifnextchar [ {\mn@doi@}
  {\mn@doi@[]}}
\def\mn@doi@[#1]#2{\def\@tempa{#1}\ifx\@tempa\@empty \href
  {http://dx.doi.org/#2} {doi:#2}\else \href {http://dx.doi.org/#2} {#1}\fi
  \endgroup}
\def\mn@eprint#1#2{\mn@eprint@#1:#2::\@nil}
\def\mn@eprint@arXiv#1{\href {http://arxiv.org/abs/#1} {{\tt arXiv:#1}}}
\def\mn@eprint@dblp#1{\href {http://dblp.uni-trier.de/rec/bibtex/#1.xml}
  {dblp:#1}}
\def\mn@eprint@#1:#2:#3:#4\@nil{\def\@tempa {#1}\def\@tempb {#2}\def\@tempc
  {#3}\ifx \@tempc \@empty \let \@tempc \@tempb \let \@tempb \@tempa \fi \ifx
  \@tempb \@empty \def\@tempb {arXiv}\fi \@ifundefined
  {mn@eprint@\@tempb}{\@tempb:\@tempc}{\expandafter \expandafter \csname
  mn@eprint@\@tempb\endcsname \expandafter{\@tempc}}}

\bibitem[\protect\citeauthoryear{{Abramowicz}, {Czerny}, {Lasota}  \&
  {Szuszkiewicz}}{{Abramowicz} et~al.}{1988}]{Abramowicz_1988}
{Abramowicz} M.~A.,  {Czerny} B.,  {Lasota} J.~P.,   {Szuszkiewicz} E.,  1988,
  \mn@doi [\apj] {10.1086/166683}, \href
  {https://ui.adsabs.harvard.edu/abs/1988ApJ...332..646A} {332, 646}

\bibitem[\protect\citeauthoryear{{Abramowicz}, {Chen}, {Kato}, {Lasota}  \&
  {Regev}}{{Abramowicz} et~al.}{1995}]{Abramowicz_1995}
{Abramowicz} M.~A.,  {Chen} X.,  {Kato} S.,  {Lasota} J.-P.,   {Regev} O.,
  1995, \mn@doi [\apjl] {10.1086/187709}, \href
  {https://ui.adsabs.harvard.edu/abs/1995ApJ...438L..37A} {438, L37}

\bibitem[\protect\citeauthoryear{{Ba{\~n}ados} et~al.,}{{Ba{\~n}ados}
  et~al.}{2018}]{Banados}
{Ba{\~n}ados} E.,  et~al., 2018, \mn@doi [\nat] {10.1038/nature25180}, \href
  {https://ui.adsabs.harvard.edu/abs/2018Natur.553..473B} {553, 473}

\bibitem[\protect\citeauthoryear{{Bachetti} et~al.,}{{Bachetti}
  et~al.}{2014}]{2014_Bachetti}
{Bachetti} M.,  et~al., 2014, \mn@doi [\nat] {10.1038/nature13791}, \href
  {https://ui.adsabs.harvard.edu/abs/2014Natur.514..202B} {514, 202}

\bibitem[\protect\citeauthoryear{{Bardeen}}{{Bardeen}}{1970}]{Bardeen_1970}
{Bardeen} J.~M.,  1970, \mn@doi [\nat] {10.1038/226064a0}, \href
  {https://ui.adsabs.harvard.edu/abs/1970Natur.226...64B} {226, 64}

\bibitem[\protect\citeauthoryear{{Barr{\`e}re}, {Guilet}, {Reboul-Salze},
  {Raynaud}  \& {Janka}}{{Barr{\`e}re} et~al.}{2022}]{NS_dynamo_2022}
{Barr{\`e}re} P.,  {Guilet} J.,  {Reboul-Salze} A.,  {Raynaud} R.,   {Janka}
  H.~T.,  2022, \mn@doi [\aap] {10.1051/0004-6361/202244172}, \href
  {https://ui.adsabs.harvard.edu/abs/2022A&A...668A..79B} {668, A79}

\bibitem[\protect\citeauthoryear{{Begelman}}{{Begelman}}{1979}]{Begelman_1978}
{Begelman} M.~C.,  1979, \mn@doi [\mnras] {10.1093/mnras/187.2.237}, \href
  {https://ui.adsabs.harvard.edu/abs/1979MNRAS.187..237B} {187, 237}

\bibitem[\protect\citeauthoryear{{Bondi}}{{Bondi}}{1952}]{Bondi_1952}
{Bondi} H.,  1952, \mn@doi [\mnras] {10.1093/mnras/112.2.195}, \href
  {https://ui.adsabs.harvard.edu/abs/1952MNRAS.112..195B} {112, 195}

\bibitem[\protect\citeauthoryear{{Briel}, {Stevance}  \& {Eldridge}}{{Briel}
  et~al.}{2023}]{Briel_2022_b}
{Briel} M.~M.,  {Stevance} H.~F.,   {Eldridge} J.~J.,  2023, \mn@doi [\mnras]
  {10.1093/mnras/stad399}, \href
  {https://ui.adsabs.harvard.edu/abs/2023MNRAS.tmp..441B} {}

\bibitem[\protect\citeauthoryear{{Chashkina}, {Lipunova}, {Abolmasov}  \&
  {Poutanen}}{{Chashkina} et~al.}{2019}]{Chashkina_2019}
{Chashkina} A.,  {Lipunova} G.,  {Abolmasov} P.,   {Poutanen} J.,  2019,
  \mn@doi [\aap] {10.1051/0004-6361/201834414}, \href
  {https://ui.adsabs.harvard.edu/abs/2019A&A...626A..18C} {626, A18}

\bibitem[\protect\citeauthoryear{{Chen}, {Abramowicz}  \& {Lasota}}{{Chen}
  et~al.}{1997}]{Chen_1997}
{Chen} X.,  {Abramowicz} M.~A.,   {Lasota} J.-P.,  1997, \mn@doi [\apj]
  {10.1086/303592}, \href
  {https://ui.adsabs.harvard.edu/abs/1997ApJ...476...61C} {476, 61}

\bibitem[\protect\citeauthoryear{{Czerny}}{{Czerny}}{2019}]{Czerny_2019}
{Czerny} B.,  2019, \mn@doi [Universe] {10.3390/universe5050131}, \href
  {https://ui.adsabs.harvard.edu/abs/2019Univ....5..131C} {5, 131}

\bibitem[\protect\citeauthoryear{{Eggenberger}, {den Hartogh}, {Buldgen},
  {Meynet}, {Salmon}  \& {Deheuvels}}{{Eggenberger}
  et~al.}{2019}]{Eggenberger_2019}
{Eggenberger} P.,  {den Hartogh} J.~W.,  {Buldgen} G.,  {Meynet} G.,  {Salmon}
  S.~J.~A.~J.,   {Deheuvels} S.,  2019, \mn@doi [\aap]
  {10.1051/0004-6361/201936348}, \href
  {https://ui.adsabs.harvard.edu/abs/2019A&A...631L...6E} {631, L6}

\bibitem[\protect\citeauthoryear{{Farmer}, {Renzo}, {de Mink}, {Marchant}  \&
  {Justham}}{{Farmer} et~al.}{2019}]{Farmer_PISNe_2019}
{Farmer} R.,  {Renzo} M.,  {de Mink} S.~E.,  {Marchant} P.,   {Justham} S.,
  2019, \mn@doi [\apj] {10.3847/1538-4357/ab518b}, \href
  {https://ui.adsabs.harvard.edu/abs/2019ApJ...887...53F} {887, 53}

\bibitem[\protect\citeauthoryear{{Fowler} \& {Hoyle}}{{Fowler} \&
  {Hoyle}}{1964}]{Fowler_and_Hoyle1964}
{Fowler} W.~A.,  {Hoyle} F.,  1964, \mn@doi [\apjs] {10.1086/190103}, \href
  {https://ui.adsabs.harvard.edu/abs/1964ApJS....9..201F} {9, 201}

\bibitem[\protect\citeauthoryear{{Frank}, {King}  \& {Raine}}{{Frank}
  et~al.}{2002}]{Frank_2002}
{Frank} J.,  {King} A.,   {Raine} D.~J.,  2002, {Accretion Power in
  Astrophysics}

\bibitem[\protect\citeauthoryear{{Fuller} \& {Ma}}{{Fuller} \&
  {Ma}}{2019}]{Fuller_Ma_2019}
{Fuller} J.,  {Ma} L.,  2019, \mn@doi [\apjl] {10.3847/2041-8213/ab339b}, \href
  {https://ui.adsabs.harvard.edu/abs/2019ApJ...881L...1F} {881, L1}

\bibitem[\protect\citeauthoryear{{Fuller}, {Piro}  \& {Jermyn}}{{Fuller}
  et~al.}{2019}]{Fuller_2019}
{Fuller} J.,  {Piro} A.~L.,   {Jermyn} A.~S.,  2019, \mn@doi [\mnras]
  {10.1093/mnras/stz514}, \href
  {https://ui.adsabs.harvard.edu/abs/2019MNRAS.485.3661F} {485, 3661}

\bibitem[\protect\citeauthoryear{{Gammie} \& {Popham}}{{Gammie} \&
  {Popham}}{1998}]{Gammie_Popham_1998}
{Gammie} C.~F.,  {Popham} R.,  1998, \mn@doi [\apj] {10.1086/305521}, \href
  {https://ui.adsabs.harvard.edu/abs/1998ApJ...498..313G} {498, 313}

\bibitem[\protect\citeauthoryear{{Gou} et~al.,}{{Gou}
  et~al.}{2011}]{CygnusX1:Gou_2011}
{Gou} L.,  et~al., 2011, \mn@doi [\apj] {10.1088/0004-637X/742/2/85}, \href
  {https://ui.adsabs.harvard.edu/abs/2011ApJ...742...85G} {742, 85}

\bibitem[\protect\citeauthoryear{{Illarionov} \& {Sunyaev}}{{Illarionov} \&
  {Sunyaev}}{1975}]{Illarionov_Sunyaev_1975}
{Illarionov} A.~F.,  {Sunyaev} R.~A.,  1975, \aap, \href
  {https://ui.adsabs.harvard.edu/abs/1975A&A....39..185I} {39, 185}

\bibitem[\protect\citeauthoryear{{Israel} et~al.,}{{Israel}
  et~al.}{2017}]{Israel_2017}
{Israel} G.~L.,  et~al., 2017, \mn@doi [Science] {10.1126/science.aai8635},
  \href {https://ui.adsabs.harvard.edu/abs/2017Sci...355..817I} {355, 817}

\bibitem[\protect\citeauthoryear{{Johnson} \& {Upton Sanderbeck}}{{Johnson} \&
  {Upton Sanderbeck}}{2022}]{Johnson_Sanderbeck_2022}
{Johnson} J.~L.,  {Upton Sanderbeck} P.~R.,  2022, \mn@doi [\apj]
  {10.3847/1538-4357/ac7b81}, \href
  {https://ui.adsabs.harvard.edu/abs/2022ApJ...934...58J} {934, 58}

\bibitem[\protect\citeauthoryear{{Jones}, {Brenneman}, {Civano}, {Lanzuisi}  \&
  {Marchesi}}{{Jones} et~al.}{2020}]{2020_Jones}
{Jones} M.,  {Brenneman} L.,  {Civano} F.,  {Lanzuisi} G.,   {Marchesi} S.,
  2020, \mn@doi [arXiv e-prints] {10.48550/arXiv.2008.08588}, \href
  {https://ui.adsabs.harvard.edu/abs/2020arXiv200808588J} {p. arXiv:2008.08588}

\bibitem[\protect\citeauthoryear{{Katz}}{{Katz}}{1977}]{Katz:1977}
{Katz} J.~I.,  1977, \mn@doi [\apj] {10.1086/155355}, \href
  {https://ui.adsabs.harvard.edu/abs/1977ApJ...215..265K} {215, 265}

\bibitem[\protect\citeauthoryear{{King}, {Pringle}  \& {Livio}}{{King}
  et~al.}{2007}]{King_2007}
{King} A.~R.,  {Pringle} J.~E.,   {Livio} M.,  2007, \mn@doi [\mnras]
  {10.1111/j.1365-2966.2007.11556.x}, \href
  {https://ui.adsabs.harvard.edu/abs/2007MNRAS.376.1740K} {376, 1740}

\bibitem[\protect\citeauthoryear{{Kroupa}}{{Kroupa}}{2001}]{Kroupa2001}
{Kroupa} P.,  2001, \mn@doi [\mnras] {10.1046/j.1365-8711.2001.04022.x}, \href
  {https://ui.adsabs.harvard.edu/abs/2001MNRAS.322..231K} {322, 231}

\bibitem[\protect\citeauthoryear{{Lasota}, {Vieira}, {Sadowski}, {Narayan}  \&
  {Abramowicz}}{{Lasota} et~al.}{2016}]{Lasota_2016}
{Lasota} J.~P.,  {Vieira} R.~S.~S.,  {Sadowski} A.,  {Narayan} R.,
  {Abramowicz} M.~A.,  2016, \mn@doi [\aap] {10.1051/0004-6361/201527636},
  \href {https://ui.adsabs.harvard.edu/abs/2016A&A...587A..13L} {587, A13}

\bibitem[\protect\citeauthoryear{{McClintock}, {Shafee}, {Narayan},
  {Remillard}, {Davis}  \& {Li}}{{McClintock} et~al.}{2006}]{McClintock_2006}
{McClintock} J.~E.,  {Shafee} R.,  {Narayan} R.,  {Remillard} R.~A.,  {Davis}
  S.~W.,   {Li} L.-X.,  2006, \mn@doi [\apj] {10.1086/508457}, \href
  {https://ui.adsabs.harvard.edu/abs/2006ApJ...652..518M} {652, 518}

\bibitem[\protect\citeauthoryear{{Narayan} \& {Yi}}{{Narayan} \&
  {Yi}}{1994}]{Narayan_Yi_1994}
{Narayan} R.,  {Yi} I.,  1994, \mn@doi [\apjl] {10.1086/187381}, \href
  {https://ui.adsabs.harvard.edu/abs/1994ApJ...428L..13N} {428, L13}

\bibitem[\protect\citeauthoryear{{Narayan} \& {Yi}}{{Narayan} \&
  {Yi}}{1995}]{Narayan_Yi_1995}
{Narayan} R.,  {Yi} I.,  1995, \mn@doi [\apj] {10.1086/176343}, \href
  {https://ui.adsabs.harvard.edu/abs/1995ApJ...452..710N} {452, 710}

\bibitem[\protect\citeauthoryear{{Narayan}, {Kato}  \& {Honma}}{{Narayan}
  et~al.}{1997}]{Narayan_1997}
{Narayan} R.,  {Kato} S.,   {Honma} F.,  1997, \mn@doi [\apj] {10.1086/303591},
  \href {https://ui.adsabs.harvard.edu/abs/1997ApJ...476...49N} {476, 49}

\bibitem[\protect\citeauthoryear{{Novikov} \& {Thorne}}{{Novikov} \&
  {Thorne}}{1973}]{Novikov_Thorne_1973}
{Novikov} I.~D.,  {Thorne} K.~S.,  1973, in Black Holes (Les Astres Occlus). pp
  343--450

\bibitem[\protect\citeauthoryear{{Ohsuga}, {Mineshige}, {Mori}  \&
  {Umemura}}{{Ohsuga} et~al.}{2002}]{Ohsuga_2002}
{Ohsuga} K.,  {Mineshige} S.,  {Mori} M.,   {Umemura} M.,  2002, \mn@doi [\apj]
  {10.1086/340798}, \href
  {https://ui.adsabs.harvard.edu/abs/2002ApJ...574..315O} {574, 315}

\bibitem[\protect\citeauthoryear{{Ohsuga}, {Mori}, {Nakamoto}  \&
  {Mineshige}}{{Ohsuga} et~al.}{2005}]{2005_Ohsuga}
{Ohsuga} K.,  {Mori} M.,  {Nakamoto} T.,   {Mineshige} S.,  2005, \mn@doi
  [\apj] {10.1086/430728}, \href
  {https://ui.adsabs.harvard.edu/abs/2005ApJ...628..368O} {628, 368}

\bibitem[\protect\citeauthoryear{{Paczy{\'n}sky} \& {Wiita}}{{Paczy{\'n}sky} \&
  {Wiita}}{1980}]{Paczynsky_Wiita_1980}
{Paczy{\'n}sky} B.,  {Wiita} P.~J.,  1980, \aap, \href
  {https://ui.adsabs.harvard.edu/abs/1980A&A....88...23P} {88, 23}

\bibitem[\protect\citeauthoryear{{Popham} \& {Gammie}}{{Popham} \&
  {Gammie}}{1998}]{Popham_Gammie_1998}
{Popham} R.,  {Gammie} C.~F.,  1998, \mn@doi [\apj] {10.1086/306054}, \href
  {https://ui.adsabs.harvard.edu/abs/1998ApJ...504..419P} {504, 419}

\bibitem[\protect\citeauthoryear{{Poutanen}, {Lipunova}, {Fabrika}, {Butkevich}
   \& {Abolmasov}}{{Poutanen} et~al.}{2007}]{Poutanen_2007}
{Poutanen} J.,  {Lipunova} G.,  {Fabrika} S.,  {Butkevich} A.~G.,   {Abolmasov}
  P.,  2007, \mn@doi [\mnras] {10.1111/j.1365-2966.2007.11668.x}, \href
  {https://ui.adsabs.harvard.edu/abs/2007MNRAS.377.1187P} {377, 1187}

\bibitem[\protect\citeauthoryear{{Sadowski}, {Bursa}, {Abramowicz},
  {Klu{\'z}niak}, {Lasota}, {Moderski}  \& {Safarzadeh}}{{Sadowski}
  et~al.}{2011}]{Sadowski_2011}
{Sadowski} A.,  {Bursa} M.,  {Abramowicz} M.,  {Klu{\'z}niak} W.,  {Lasota}
  J.~P.,  {Moderski} R.,   {Safarzadeh} M.,  2011, \mn@doi [\aap]
  {10.1051/0004-6361/201116702}, \href
  {https://ui.adsabs.harvard.edu/abs/2011A&A...532A..41S} {532, A41}

\bibitem[\protect\citeauthoryear{{Shakura} \& {Sunyaev}}{{Shakura} \&
  {Sunyaev}}{1973}]{Shakura_Sunyaev_1973}
{Shakura} N.~I.,  {Sunyaev} R.~A.,  1973, \aap, \href
  {https://ui.adsabs.harvard.edu/abs/1973A&A....24..337S} {24, 337}

\bibitem[\protect\citeauthoryear{{S{\k{a}}dowski} \&
  {Narayan}}{{S{\k{a}}dowski} \& {Narayan}}{2016}]{2016_Sadowski}
{S{\k{a}}dowski} A.,  {Narayan} R.,  2016, \mn@doi [\mnras]
  {10.1093/mnras/stv2941}, \href
  {https://ui.adsabs.harvard.edu/abs/2016MNRAS.456.3929S} {456, 3929}

\bibitem[\protect\citeauthoryear{{S{\k{a}}dowski}, {Narayan}, {McKinney}  \&
  {Tchekhovskoy}}{{S{\k{a}}dowski} et~al.}{2014}]{2014_Sadowski}
{S{\k{a}}dowski} A.,  {Narayan} R.,  {McKinney} J.~C.,   {Tchekhovskoy} A.,
  2014, \mn@doi [\mnras] {10.1093/mnras/stt2479}, \href
  {https://ui.adsabs.harvard.edu/abs/2014MNRAS.439..503S} {439, 503}

\bibitem[\protect\citeauthoryear{{Thorne}}{{Thorne}}{1974}]{Thorne_1974}
{Thorne} K.~S.,  1974, \mn@doi [\apj] {10.1086/152991}, \href
  {https://ui.adsabs.harvard.edu/abs/1974ApJ...191..507T} {191, 507}

\bibitem[\protect\citeauthoryear{{Trakhtenbrot}, {Volonteri}  \&
  {Natarajan}}{{Trakhtenbrot} et~al.}{2017}]{2017_Trakhtenbrot}
{Trakhtenbrot} B.,  {Volonteri} M.,   {Natarajan} P.,  2017, \mn@doi [\apjl]
  {10.3847/2041-8213/836/1/L1}, \href
  {https://ui.adsabs.harvard.edu/abs/2017ApJ...836L...1T} {836, L1}

\bibitem[\protect\citeauthoryear{{Volonteri}, {Habouzit}  \&
  {Colpi}}{{Volonteri} et~al.}{2021}]{Volonteri_2021}
{Volonteri} M.,  {Habouzit} M.,   {Colpi} M.,  2021, \mn@doi [Nature Reviews
  Physics] {10.1038/s42254-021-00364-9}, \href
  {https://ui.adsabs.harvard.edu/abs/2021NatRP...3..732V} {3, 732}

\bibitem[\protect\citeauthoryear{{van Son} et~al.,}{{van Son}
  et~al.}{2020}]{van_son_2020}
{van Son} L.~A.~C.,  et~al., 2020, \mn@doi [\apj] {10.3847/1538-4357/ab9809},
  \href {https://ui.adsabs.harvard.edu/abs/2020ApJ...897..100V} {897, 100}

\makeatother
\end{thebibliography}

\appendix

\vspace{-0.6cm}
\section{Approximating radial velocity} \label{sec: derive g/f_*}

Let us assume that $g/f_*$ takes the form 
\begin{equation}
   g/f_* = (r/r_g)^{\frac{1}{2} h\left(  \frac{r}{r_g}\right)},
\end{equation} 
where $h$ is some unknown function of $r/r_g$. On substituting this into Eq. \ref{eq: radial velocity}, the resulting expression could be rewritten as
\begin{equation}
    {\rm log} \frac{{\rm v}_r}{c} = {\rm log}  \left(\frac{3 \alpha \lambda^2}{2^{3/2} }\frac{r}{r - r_g} \right)   -  \frac{1-h}{2}  {\rm log} \frac{r}{r_g}.
    \label{eq: log of v_r}
\end{equation}
It can be seen that the slope of ${\rm v}_r$ in Fig. \ref{fig: extended solution} would be determined by the coefficient of the second logarithm on RHS above. As we can already calculate the value of $g/f_*$ at $r = r_s$ and $10^4 r_g$, this means $h$ is also known at these boundary points, since 
\begin{equation}
    h = \frac{2 \ln{g/f_*}}{\ln{r/r_g}}.
\end{equation}
As the slope of ${\rm v}_r$ is inversely proportional to $r$, hence a simple guess shows that for our purpose $h \propto r^{-3/2}$ fairly approximates the form of ${\rm v}_r$ between the two boundary points.






\bsp	
\label{lastpage}
\end{document}